\documentclass[12pt]{iopart}
\usepackage{amsmath,amsfonts,amssymb,bbm,citesort,graphicx}

\newcommand{\eqn}[1]{\begin{equation} #1 \end{equation}} % equation environment
\newcommand{\aln}[1]{\begin{align} #1 \end{align}}       % align environment
\newcommand{\mc}{\mathcal}                               % script letters
\newcommand{\msf}{\mathsf}                               % straight letters
\newcommand{\mbf}{\mathbf}                               % bold letters
\newcommand{\eq}[1]{(\ref{#1})}              % equation reference
\newcommand{\pd}{\partial}                   % partial derivative sign
\newcommand{\wtilde}{\widetilde}             % wide tilde
\newcommand{\wbar}{\overline}                % wide bar
\renewcommand{\l}{\left}                     % left bracket
\renewcommand{\r}{\right}                    % right bracket

\begin{document}

\title{Gas lasers with wave-chaotic resonators}

\author{Oleg Zaitsev}
\ead{oleg.zaitsev@uni-bonn.de}
\address{Physikalisches Institut der Universit\"at Bonn, Nu{\ss}allee 12,
             53115 Bonn, Germany}

\begin{abstract}

Semiclassical multimode laser theory is extended to gas lasers with open
two-dimensional resonators of arbitrary shape.  The Doppler frequency shift of
the linear-gain coefficient leads to an additional linear coupling between the
modes, which, however, is shown to be negligible.  The nonlinear laser equations
simplify in the special case of wave-chaotic resonators. In the single-mode
regime, the intensity of a chaotic laser, as a function of the mode frequency,
displays a local minimum at the frequency of the atomic transition. The width of
the minimum scales with the inhomogeneous linewidth, in contrast to the Lamb dip
in uniaxial resonators whose width is given by the homogeneous linewidth.

\end{abstract}

%\submitto{\jpb}
%\maketitle

\section{Introduction}

Theory of gas lasers was developed some decades ago and is well covered in the
literature~\cite{sarg74,hake84}. The gas lasers are characterized by their
capability to emit light at frequencies that differ from the atomic-transition
frequency by more than a homogeneous linewidth. This property is a consequence
of the coherent amplification by moving atoms that interact with the electric
field at Doppler-shifted frequencies. If a resonance frequency of the cavity is
continuously changed, the intensity of the single mode drops within the
homogeneous linewidth of the atomic frequency. This narrow local minimum in the
inhomogeneously broadened line, the Lamb dip~\cite{lamb64}, can be observed in
uniaxial cavities that support standing waves. The Lamb dip has been utilized
for stabilization of laser frequency~\cite{free70} and precise detection of
atomic-transition frequency and line shape (Lamb-dip
spectroscopy)~\cite{szok66}. 

The theory of lasers (including gas lasers) was originally intended for low-loss
resonators with a well-defined direction of propagation. The continuing interest
in random~\cite{cao99,pols01b} and chaotic~\cite{misi98} lasers has motivated
extensions of the conventional theory to irregular and strongly open
systems~\cite{misi98,patr00,vivi03,deyc05a,ture06,zait10b}. These improvements
relate to a number of aspects, such as construction of a quasimode basis of the
open cavity, description of linear and nonlinear coupling between the lasing
modes in irregular lasers, and incorporating the statistics of eigenfrequencies
and eigenfunctions in the laser equations (see Ref.~\cite{zait10} for a review).

In the present article, the semiclassical theory of gas lasers is generalized
to lasers with open two-dimensional resonators of arbitrary shape. Equations
for the amplitudes and frequencies of lasing modes are derived by expanding the
atomic polarization in the powers of electric field and keeping the linear and
third-order contributions.  Since the dependence of the linear gain on the
frequency is modified by the Doppler shifts in the frames of moving atoms,
there exists an additional linear coupling between the modes. This effect is
similar to the gain-induced coupling in systems with spatially inhomogeneous
refractive index~\cite{deyc05a} or pumping~\cite{zait10}. We can argue,
however, that the the magnitude of the coupling in gas lasers is negligible
under reasonable assumptions. The nonlinear terms in the laser equations
simplify for the weakly open cavities of chaotic geometry, where the eigenmodes
can be modeled by a superposition of plane waves propagating in random
directions~\cite{berr77,berr83}. In this system, the intensity of a single
lasing mode, as a function of its frequency, has a local minimum at the
frequency of the atomic transition. In contrast to the Lamb dip in uniaxial
resonators, the width of the minimum is comparable to the inhomogeneous
linewidth.

\section{Semiclassical theory of gas lasers}

\subsection{Semiclassical laser equations}

We consider an open resonator of arbitrary shape, which, in general, can be
filled with a homogeneous inactive medium characterized by a real dielectric
constant~$\epsilon$. The lasing action is provided by active atoms located
inside the resonator. In the simplest model, only the the two atomic levels
responsible for the lasing transition are described dynamically, whereas the
information about the remaining degrees of freedom is contained in the
empirical parameters: the homogeneous atomic linewidth~$\gamma_\perp$, the
population-inversion decay rate~$\gamma_\parallel$, and the unsaturated
population difference (per unit volume)~$\Delta n_0$. Then the system of atoms
can be characterized by the polarization~$\mbf P (\mbf r, t)$ and the
population-inversion density (the difference between the populations of the
upper and lower levels per volume)~$\Delta n (\mbf r, t)$. The interaction of
these quantities and the classical electric field~$\mbf E (\mbf r, t)$ in the
resonator is specified by the semiclassical laser equations~\cite{hake84}
\aln{
  &\frac \epsilon {c^2} \frac {\pd^2 \mbf E} {\pd t^2} + \nabla \times (\nabla
  \times \mbf E) = -\frac {4 \pi} {c^2} \frac {\pd^2} {\pd t^2} \mbf P (\mbf r,
  t), 
\label{E}\\
  &\l(\frac {\pd^2} {\pd t^2} + 2 \gamma_\perp \frac \pd {\pd t} + \nu^2 \r)
  \mbf P = -2 \nu \frac {d^2} \hbar \mbf E (\mbf r, t)\, \Delta n (\mbf r, t),
\label{P}\\
  &\frac {\pd \Delta n} {\pd t} - \gamma_\parallel (\Delta n_0 - \Delta n) =
  \frac 2 {\hbar \nu} \mbf E (\mbf r, t) \cdot \frac {\pd \mbf P (\mbf r, t)}
  {\pd t},
\label{Deltan}
}
where $\nu$ is the atomic-transition frequency and $d$ is the atomic dipole
matrix element. The first equation follows from the Maxwell equations (written
in the Gaussian units) and has the form of a wave equation with the polarization
as a source. The second equation describes the dipole interaction with the field
and the decay of polarization due to stochastic degrees of freedom. The third
equation describes the energy exchange between the atoms and the field, as well
as the relaxation of the population inversion. If the field were absent, $\Delta
n (\mbf r, t)$~would reach the stationary value of~$\Delta n_0$. We will use
this parameter as a measure of the pump strength. 

In the gas laser, the active atoms are moving with constant velocities~$\mbf
v_j$, which are assumed to be Maxwell distributed. The polarization and the
population-inversion density can be expressed in terms of the dipole
moments~$\mbf p_j (t)$ and the population differences~$\Delta N_j (t)$ of the
individual atoms~as
\aln{
  &\mbf P (\mbf r, t) = \sum_{j = 1}^{\mc N} \mbf p_j (t)\, \delta \bigl(\mbf r
  - \mbf r_j (t) \bigr),
\label{pj}\\
  &\Delta n (\mbf r, t) = \sum_{j = 1}^{\mc N} \Delta N_j (t)\, \delta
  \bigl(\mbf r - \mbf r_j (t) \bigr),
\label{DeltaNj}
}
where $\mc N$ is the number of atoms and $\mbf r_j (t) = \mbf r_j^0 + \mbf v_j
t$ is the trajectory of the $j$th atom. The initial positions $\mbf r_j^0$ are
uniformly distributed within the resonator. The change of direction due to
collisions or reflections at the boundary can be neglected if the time between
these events is longer than the longest time scale in Eqs.~\eq{E}-\eq{Deltan},
i.e., $l, L \gg \wbar v / \gamma_\parallel$, where $l$ is the mean free path,
$L$~is the size of the system, and $\wbar v$~is the root mean square velocity. 

We will consider the case of a quasi-two-dimensional resonator, and assume that
the field vector $\mbf E (\mbf r, t)$ has a fixed polarization direction
perpendicular to the plane of the resonator. This assumption is consistent with
the Coulomb gauge condition $\nabla \cdot [\epsilon \mbf E (\mbf r, t)] = 0$.
Then the electric field and the dipole moments can be described in terms of
their out-of-plane components, $E (\mbf r, t)$ and~$p_j (t)$.

Clearly, Eqs.~\eq{P} and~\eq{Deltan} are local in the sense that the position
$\mbf r$ enters only as a parameter. Hence, the contributions of individual
atoms can be separated by substituting the sums~\eq{pj} and~\eq{DeltaNj} and
collecting the terms proportional to the respective delta functions:
\aln{
  &\l(\frac {\pd^2} {\pd t^2} + 2 \gamma_\perp \frac \pd {\pd t} + \nu^2 \r) p_j
  = -2 \nu \frac {d^2} \hbar E \bigl(\mbf r_j (t), t \bigr)\, \Delta N_j (t), 
\label{pj_eq}\\
  &\frac {d \Delta N_j} {d t} - \gamma_\parallel (\Delta N_0 - \Delta N_j) 
  = \frac 2 {\hbar \nu} E \bigl(\mbf r_j (t), t \bigr)\, \frac {d
  p_j (t)} {d t},
\label{DeltaNj_eq}
}
where $\Delta N_0 = \Delta n_0 V/ \mc N$ is the unsaturated inversion per atom
and $V$ is the resonator volume. 

Since the field in Eqs.~\eq{pj_eq} and~\eq{DeltaNj_eq} is evaluated at the
positions of moving atoms, it is convenient to perform the Fourier transform 
\aln{
  &E (\mbf r, t) = \frac 1 {(2 \pi)^2} \int d \mbf q\, E_{\mbf q} (t)\, e^{i
  \mbf q \cdot \mbf r}, \\
  &E_{\mbf q} (t) = \int_{\mc I} d \mbf r\, E (\mbf r, t)\, e^{-i \mbf q \cdot
  \mbf r},
\label{Eq}
}
where the vectors $\mbf r$ and $\mbf q$ are two-dimensional. The transformation
is applied only to the field inside the cavity, the integration being performed
over the cavity interior~$\mc I$. In the moving frame of the atom, the Fourier
component $E_{\mbf q} (t)$ is multiplied by an additional oscillating factor
$\exp[i \mbf q \cdot \mbf v_j t]$, which expresses the Doppler shift for the
frequency. In order to facilitate subsequent handling of lasing field
oscillations, we carry out the temporal Fourier transform as well: 
\aln{
  &\l\{\begin{array}{c}
    E_{\mbf q} (t) \\
    p_j (t) \\
    \Delta N_j (t)
  \end{array}\r\}
  = \frac 1 {2 \pi} \int_{-\infty}^\infty d \omega
  \l\{\begin{array}{c}
    E_{\mbf q \omega} \\
    p_{j \omega} \\
    \Delta N_{j \omega} 
  \end{array}\r\}
  e^{- i \omega t}, \\
  &\l\{\begin{array}{c}
    E_{\mbf q \omega} \\
    p_{j \omega} \\
    \Delta N_{j \omega} 
  \end{array}\r\}
  = \int_{-\infty}^\infty dt 
  \l\{\begin{array}{c}
    E_{\mbf q} (t) \\
    p_j (t) \\
    \Delta N_j (t)
  \end{array}\r\}
  e^{i \omega t}.
}

In the $(\mbf q, \omega)$ representation, the atom equations \eq{pj_eq}
and~\eq{DeltaNj_eq} assume the form:
\aln{
  &\l(- \omega^2 - 2i \gamma_\perp \omega + \nu^2 \r) p_{j \omega}  \notag \\
  &= -2 \nu \frac {d^2} \hbar \frac 1 {(2 \pi)^3} \int d \mbf q\, d \omega' \,
  e^{i \mbf q \cdot \mbf r_j^0}\, E_{\mbf q \omega'} \Delta N_{j, \omega -
  \omega' + \mbf q \cdot \mbf v_j}, 
\label{pom}\\
  &(- i \omega + \gamma_\parallel) \Delta N_{j \omega} = 2 \pi \gamma_\parallel
  \Delta N_0\, \delta (\omega) \notag \\
  &- \frac {2i} {\hbar \nu} \frac 1 {(2 \pi)^3} \int d \mbf q\, d \omega'
  \omega' e^{i \mbf q \cdot \mbf r_j^0}\, p_{j \omega'} E_{\mbf q, \omega -
  \omega' + \mbf q \cdot \mbf v_j}.
\label{deltaNom}
}
Together with the wave equation~\eq{E}, the above equations form a closed
system, which, in principle, allows one to determine the electric field in the
cavity.

\subsection{Third-order perturbation expansion}

In order to derive an equation for the electric field alone, the dipole moments
$p_{j \omega}$ can be expressed in terms of the field using Eqs.~\eq{pom}
and~\eq{deltaNom} and then substituted in the polarization term of the wave
equation~\eq{E}. To eliminate the inversion variable~$\Delta N_{j \omega}$ from
the atom equations, we expand $p_{j \omega}$ and $\Delta N_{j \omega}$ in
powers of the field amplitudes~$E_{\mbf q \omega}$. Here, the perturbation
expansion is carried out up to the third-order terms, which provides the
minimal description of the nonlinear effects and is a standard procedure in the
laser theory~\cite{sarg74,hake84}. The third-order theory becomes invalid if
the mode intensity is far above the lasing threshold, and a fully nonlinear
approach along the lines of Refs.~\cite{ture06,zait10b} has to be developed. 
The nonperturbative theories rely on the possibility to group terms according
to their order of magnitude in the small parameter $\gamma_\parallel  /\delta
\omega$, where $\delta \omega$ is the frequency spacing between the lasing
modes. These contributions can be traced back to oscillations of the population
inversion, the zeroth order corresponding to the stationary-inversion
approximation of Ref.~\cite{ture06}. The oscillations of the population at the
beat frequencies in the multimode regime yield a correction to the polarization
(linear in $\gamma_\parallel  /\delta \omega$ and third order in the field),
which oscillates at the same lasing frequency as the main contribution (zeroth
order in $\gamma_\parallel  /\delta \omega$) to the
polarization~\cite{sarg74,zait07}. This is the level of approximation used in
the present section. The leading-order corrections in $\gamma_\parallel 
/\delta \omega$ to the fully nonlinear theory of Ref.~\cite{ture06} were
derived and numerically studied in Ref.~\cite{ge08}. A diagrammatic expansion
in $\gamma_\parallel  /\delta \omega$, nonperturbative in the field, was
proposed in Ref.~\cite{zait10b}. In the frequency representation, the inversion
pulsations are described by the function $D_\parallel (\omega)$ defined below.
In the laser with gaseous active medium, the Doppler frequency shifts mix the
orders of magnitude of the pulsation terms, and additional procedures have to
be introduced in order to approximately reclassify them. In
Sec.~\ref{sec:nlin}, the stationary-inversion approximation is obtained for the
nonlinear contribution of the third order in the electric field. 

We start with the zeroth-order approximation to the population inversion by
setting it equal to the unsaturated value: $\Delta N_{j \omega}^{(0)} = 2 \pi
\Delta N_0\, \delta (\omega)$. Substitution in Eq.~\eq{pom} yields the linear
approximation to the dipole moment,
\eqn{
  p_{j \omega}^{(1)} = -i \frac {d^2} {\hbar \gamma_\perp} \Delta N_0 \wtilde D
  (\omega) \frac 1 {(2 \pi)^2} \int d \mbf q' e^{i \mbf q' \cdot \mbf r_j^0}\,
  E_{\mbf q', \omega + \mbf q' \cdot \mbf v_j},
\label{p1}
}
where we introduced the notation 
\eqn{
  \wtilde D (\omega) = - \frac {2 i \gamma_\perp \nu} {\nu^2 - \omega^2 - 2i
  \gamma_\perp \omega}.
}
The real part of $\wtilde D (\omega)$ provides the frequency dependence of the
linear gain for atoms at rest.  By substituting $p_{j \omega}^{(1)}$ in
Eq.~\eq{deltaNom}, we obtain the quadratic correction to the inversion, $\Delta
N_{j \omega}^{(2)}$, which is substituted back into Eq.~\eq{pom} to produce the
nonlinear contribution,
\aln{
  p_{j \omega}^{(3)} = &\frac {2 i} {\hbar \nu \gamma_\parallel} \l(\frac {d^2}
  {\hbar \gamma_\perp} \r)^2 \Delta N_0\, \wtilde D (\omega)  \frac 1 {(2
  \pi)^8} \int d \mbf q' d \mbf q'' d \mbf q''' d \omega' d \omega'' \omega'
  \wtilde D (\omega')\, D_\parallel (\omega - \omega'' + \mbf q''' \cdot \mbf
  v_j) \notag \\
  &\times e^{i (\mbf q' + \mbf q'' + \mbf q''') \cdot \mbf r_j^0} E_{\mbf q'\!,
  \omega - \omega' - \omega'' + (\mbf q' + \mbf q''') \cdot \mbf v_j}  E_{\mbf
  q''\!, \omega' + \mbf q'' \cdot \mbf v_j} E_{\mbf q''' \omega''},
\label{p3}
}
where 
\eqn{
   D_\parallel (\omega) = \l( 1 - i \frac \omega {\gamma_\parallel} \r)^{-1}.
}

The resulting expansion for the dipole moment, $p_{j \omega} = p_{j
\omega}^{(1)} + p_{j \omega}^{(3)}$, serves as a source in Eq.~\eq{E}, which,
after taking into account Eq.~\eq{pj} and transformation to the
$\omega$~representation, takes the form
\aln{
  &\l(\epsilon \omega^2 + c^2 \nabla^2\r) E_{\omega} (\mbf r)  \notag \\
  &= - 4 \pi \sum_{j = 1}^{\mc N} \frac 1 {(2 \pi)^2} \int d \mbf q\, e^{i \mbf
  q \cdot (\mbf r - \mbf r_j^0)} (\omega - \mbf q \cdot \mbf v_j)^2\, p_{j,
  \omega - \mbf q \cdot \mbf v_j}. 
\label{Eom}
}
A possible means to solve this differential equation is to search for the
solution $E_{\omega} (\mbf r)$ in the form of an expansion in the eigenfunctions
of the differential operator. In the case of an open cavity, one can either
employ the continuous basis of eigenmodes of the infinite system or formulate a
non-Hermitian eigenvalue problem by separating the functional spaces of the
interior and exterior of the cavity. In the latter case, the complex
eigenfrequencies provide the oscillation frequencies and lifetimes of the
decaying quasimodes, whereas the right eigenfunctions yield their field
distributions. Here we make use of the quasimode basis, as the quasimodes
supply, in most cases, the correct description of lasing modes near the lasing
threshold. 

Specifically, we employ the constant-flux (CF) quasimodes~\cite{ture06}, which
are particularly suitable for laser problems (see \ref{sec:cf} for more
information). The CF modes are characterized by smooth wavefunctions $\psi_{k
\omega} (\mbf r)$ depending on the real continuous parameter~$\omega$. In the
asymptotic region outside the cavity, $\psi_{k \omega} (\mbf r) \propto \exp (i
\omega r/c)$~describes a wave with frequency~$|\omega|$ carrying constant energy
flux. Within the cavity, the function satisfies the Helmholtz equation with the
complex wavenumber $\Omega_k (\omega)/c$. For a given~$\omega$, there exists an
infinite set of eigenmodes labeled by the discrete (multi-)index~$k$. The
functions $\psi_{k \omega} (\mbf r)$ are the right eigenfunctions of a certain
non-Hermitian operator. The conjugate left eigenfunctions, $\phi_{k \omega}
(\mbf r)$, behave as $\exp (-i \omega r/c)$ in the asymptotic region and have
eigenfrequencies $[\Omega_k (\omega)]^*$ inside the cavity. The CF
eigenfunctions $\psi_{k \omega} (\mbf r)$~and $\phi_{k \omega} (\mbf r)$ form
complete biorthogonal bases of the open cavity. 

Let us represent  the electric field inside the cavity as a superposition of
eigenfunctions~$\psi_{k \omega} (\mbf r)$. After substitution of the
expansion~\eq{normmod} in the left-hand side of Eq.~\eq{Eom}, application of
differential equation~\eq{cfin} and biorthogonality relation~\eq{biorth}, we
obtain an equation for the amplitude $a_{k \omega}$ of the mode~$k$,
\aln{
  &\l[\Omega_k^2 (\omega) - \omega^2 \r] a_{k \omega} = \notag \\
  &\frac {4 \pi} {\sqrt \epsilon} \sum_{j = 1}^{\mc N} \frac 1 {(2 \pi)^2} \int
  d \mbf q\, e^{- i \mbf q \cdot \mbf r_j^0} (\omega - \mbf q \cdot \mbf
  v_j)^2 \phi_{k \mbf q \omega}^*\, p_{j, \omega - \mbf q \cdot \mbf v_j},
\label{ampl}
}
where $\phi_{k \mbf q \omega}$ is the Fourier transform of $\phi_{k \omega}
(\mbf r)$, defined as in Eq.~\eq{Eq}. The sum $\sum_{j = 1}^{\mc N}$ can be
substituted by the average $\mc N \langle \cdots \rangle$ over the positions
$\mbf r_j^0$ and velocities $\mbf v_j$ of active atoms. The distributions of the
positions and velocities are uncorrelated: the former are distributed uniformly
within the cavity and the latter obey the Maxwell distribution. We approximate
\eqn{
  \l< e^{i \mbf q \cdot \mbf r_j^0} \r>_{\mbf r_j^0} \simeq \frac 1 A
  \int_{\mc I} d \mbf r\, e^{i \mbf q \cdot \mbf r} = \frac {(2 \pi)^2} A
  \delta_{\mc I} (\mbf q),
}
where $(2 \pi)^2 [\delta_{\mc I} (\mbf q)]^*$ is the Fourier transform of the
characteristic function of the cavity interior and $A$~is the cavity area.
$\delta_{\mc I} (\mbf q)$~has a $\delta$-function property $\int d \mbf q\,
\delta_{\mc I} (\mbf q)\, f_{\mbf q} = f_{\mbf q = 0}$, if $f_{\mbf q}$ is the
Fourier transform of a function $f (\mbf r)$ that vanishes outside of the
cavity. By substituting the linear and third-order contributions to the dipole
moment, Eqs.~\eq{p1} and~\eq{p3}, we transform Eq.~\eq{ampl}~to 
\eqn{
  \l[\Omega_k^2 (\omega) - \omega^2 \r] a_{k \omega} = \mc P_{k \omega}^{(1)} +
  \mc P_{k \omega}^{(3)},
\label{ampl_av}
}
where the linear and third-order terms are, respectively,
\aln{
  \mc P_{k \omega}^{(1)} =  &- i c^{(1)} \frac 1 {(2 \pi)^2} \int d \mbf
  q\, \phi_{k \mbf q \omega}^*  E_{\mbf q \omega} \notag \\ 
  &\times \l<  (\omega - \mbf q \cdot \mbf v)^2 \wtilde D (\omega - \mbf q \cdot
  \mbf v) \r>_{\mbf v} , 
\label{P1}
}
\aln{
  &\mc P_{k \omega}^{(3)} =  i c^{(3)} \frac 1 {(2 \pi)^8} \int d \mbf q\, d
  \mbf q' d \mbf q'' d \omega' d \omega'' \omega' \wtilde D (\omega')\, \phi_{k
  \mbf q \omega}^*  E_{\mbf q - \mbf q' - \mbf q''\!, \omega''} \notag \\
  &\times \l< (\omega - \mbf q \cdot \mbf v)^2\, \wtilde D (\omega - \mbf q
  \cdot \mbf v)\, D_\parallel [\omega - \omega'' - (\mbf q' + \mbf q'') \cdot
  \mbf v]\, E_{\mbf q'\!, \omega - \omega' - \omega'' - \mbf q'' \cdot \mbf v}\,
  E_{\mbf q''\!, \omega' + \mbf q'' \cdot \mbf v} \r>_{\mbf v}.
\label{P3}
}
In the velocity averages we dropped the index~$j$ labeling individual atoms. The
coefficients are given~by 
\aln{
  &c^{(1)} = 4 \pi\frac {\Delta N_0\, \mc N} {\sqrt \epsilon A} \frac {d^2}
  {\hbar \gamma_\perp}, \\
  &c^{(3)} = 8 \pi \frac {\Delta N_0\, \mc N} {\sqrt \epsilon A} \frac 1 {\hbar
  \nu \gamma_\parallel} \l( \frac {d^2} {\hbar \gamma_\perp} \r)^2.
}
The field amplitudes $E_{\mbf q \omega}$ in Eqs.~\eq{P1} and~\eq{P3} can be
expanded in the CF quasimodes as in Eq.~\eq{normmod}. Therefore,
Eq.~\eq{ampl_av}~represents a system of algebraic equations for the quasimode
amplitudes~$a_{k \omega}$.

\subsection{Lasing modes}

The frequency integrals in Eq.~\eq{P3} can be eliminated by approximating the
electric field in the time domain with a superposition of constant-frequency
oscillations, the lasing modes. It is well known that when the pumping increases
above a certain threshold value, the first lasing mode appears. For higher
pumping levels, additional modes start oscillating. In the multimode regime,
the procedure of representing the field spectrum as a finite number of
frequencies is an approximation: to preserve self-consistency of the laser
equations, the oscillations at beat frequencies have to be neglected.

We start with an ansatz
\eqn{
  E_{\mbf q \omega} = 2 \pi \sum_{l = 1}^{N_{\text m}} \l[ E_{l \mbf q}\, \delta
  (\omega - \omega_l) + E_{l, - \mbf q}^*\, \delta (\omega + \omega_l) \r],
\label{Emod}
}
which assumes the existence of $N_{\text m}$ lasing modes with frequencies
$\omega_l > 0$ and guarantees the realness of~$E (\mbf r, t)$. By expanding
$E_{l \mbf q}$ in the CF quasimodes~as
\eqn{
  E_{l \mbf q} = \epsilon^{-1/2} \sum_k a_{lk} \psi_{k \mbf q \omega_l},
\label{lmqm}
}
we obtain the lasing-mode representation for the coefficients~$a_{k \omega}$,
\eqn{
  a_{k \omega} = 2 \pi \sum_{l = 1}^{N_{\text m}} \l[ a_{lk}\, \delta (\omega -
  \omega_l) + a_{lk}^*\, \delta (\omega + \omega_l) \r].
\label{amod}
}
Usually, the quasimode with the longest lifetime starts oscillating at the
lasing threshold. At higher pumping levels, the nonlinear terms in the laser
equations mix the quasimodes, and the lasing mode no longer coincides with a
single quasimode. Thus, the expansion~\eq{lmqm} is nontrivial, as it contains,
in general, more than one term.

After substituting the mode representations~\eq{Emod} and~\eq{amod} in the laser
equations~\eq{ampl_av} and collecting the terms proportional to $\delta (\omega
- \omega_l)$ [or, equivalently, $\delta (\omega + \omega_l)$], we would obtain a
rather cumbersome equation for the coefficients~$a_{lk}$. We write this equation
in a simplified form by applying the standard approximation $\delta \omega \ll
\nu$, where $\delta \omega$ is the frequency spacing between the lasing modes.
This approximation corresponds to neglecting the second time derivatives in
Eqs.~\eq{E} and~\eq{P}. Additionally, we neglect the terms containing
$D_\parallel (2 \nu) \sim \gamma_\parallel / \nu \ll \delta \omega / \nu$. The
resulting equation reads
\eqn{
  (\Omega_k - \omega_l)\, a_{lk} = \mc P_{lk}^{(1)} + \mc P_{lk}^{(3)},
\label{mod_ampl}
}
with
\eqn{
  \mc P_{lk}^{(1)} = - \frac i  2 c^{(1)} \nu \frac 1 {(2 \pi)^2} \int d \mbf
  q\, \phi_{k \mbf q}^*  \l<  D (\omega_l - \mbf q \cdot \mbf v) \r>_{\mbf v}
  E_{l \mbf q}, 
\label{Plk1} 
}
\aln{
  &\mc P_{lk}^{(3)} = \frac i 2 c^{(3)} \nu^2 \sum_{l' = 1}^{N_{\text m}}
  \frac 1 {(2 \pi)^6} \int d \mbf q\, d \mbf q' d \mbf q'' \phi_{k \mbf q}^*
  E_{l' \mbf q'}^* \notag \\
  &\times \l\{ \l< D (\omega_l - \mbf q \cdot \mbf v) \l[D (\omega_{l'}
  - \mbf q'' \cdot \mbf v) + D^* (\omega_{l'} - \mbf q' \cdot \mbf v) \r] 
  D_\parallel [(\mbf q' - \mbf q'') \cdot \mbf v] \r>_{\mbf v} E_{l, \mbf q +
  \mbf q' - \mbf q''} E_{l' \mbf q''}  \r.  \notag \\  
  &+ (1 - \delta_{ll'}) \langle D (\omega_l - \mbf q \cdot \mbf v)  \l[D
  (\omega_l - \mbf q'' \cdot \mbf v) + D^* (\omega_{l'} - \mbf q' \cdot \mbf v)
  \r]   \notag \\
  &\times \l. D_\parallel [\omega_l - \omega_{l'} + (\mbf q' - \mbf q'') \cdot
  \mbf v]  \rangle_{\mbf v}  E_{l', \mbf q + \mbf q' - \mbf q''} E_{l \mbf q''}
  \r\}.
\label{Plk3}
}
In the equations above we introduced the notation $\Omega_k \equiv \Omega_k
(\nu)$, $\phi_{k \mbf q} \equiv \phi_{k \mbf q \nu}$, and 
\eqn{
  D (\omega) = \l( 1 - i \frac {\omega - \nu} {\gamma_\perp} \r)^{-1},
\label{Dom}
}
which approximates $\wtilde D (\omega)$ for positive frequencies.  The field
amplitudes $E_{l \mbf q}$ are assumed to be expanded in the CF wavefunctions
$\psi_{k \mbf q} \equiv \psi_{k \mbf q \nu}$, as in Eq.~\eq{lmqm}. The nonlinear
laser equations~\eq{mod_ampl} can be solved for the frequencies and amplitudes
of the lasing modes, e.g., by using the iteration procedure described in
Ref.~\cite{ture08}. 

The term proportional to $(1 - \delta_{ll'})$ in Eq.~\eq{Plk3} is usually
omitted in the case of fixed atoms, $\mbf v = 0$. In such a system, the factor
$D_\parallel (\omega_l - \omega_{l'}) \sim \gamma_\parallel  /\delta \omega 
\sim \gamma_\parallel / \gamma_\perp  \ll 1$, arising due to oscillations of the
population inversion, makes this term small compared to the first term in the
braces. In Sec.~\ref{sec:nlin} we argue that for a gas laser the second term is
negligible as well.

Although the algebraic equation~\eq{mod_ampl} has been derived above for the
steady-state lasing oscillations, $a_{lk} \exp (- i \omega_l t)$, it can be
easily generalized~\cite{zait10} to the case of slowly-varying amplitudes
$a_{lk} (t)$ by adding the term $- i \dot a_{lk}$ to the left-hand side. The
resulting first-order nonlinear differential equation can be used to study
stability of the lasing modes.

\section{Properties of the laser equations}

In this section, we simplify the right-hand side of the laser
equations~\eq{mod_ampl} by taking into account the resonant dependence of the
quasimode wavefunctions on~$q$. In particular, we show that the linear
gain-induced coupling of quasimodes (Sec.~\ref{sec:lcq}) and the nonlinear
contribution proportional to $(1 - \delta_{ll'})$ (Sec.~\ref{sec:nlin}) are
negligible.

\subsection{Linear coupling of quasimodes}
\label{sec:lcq}

After the field amplitude $E_{l \mbf q}$ is expanded in the basis functions
$\psi_{k \mbf q}$, the linear contribution~\eq{Plk1} can be written~as 
\eqn{
  \mc P_{lk}^{(1)} = - \frac {i \nu} {2 \sqrt \epsilon}\, c^{(1)} \sum_{k'}
  V_{kk'} (\omega_l)\, a_{lk'}.
\label{Plk1exp}
}
The matrix elements
\eqn{
  V_{kk'} (\omega_l) = \frac 1 {(2 \pi)^2} \int d \mbf q\, \phi_{k \mbf q}^* 
  \l<  D (\omega_l - \mbf q \cdot \mbf v) \r>_{\mbf v} \psi_{k' \mbf q}
\label{Vlkk}
}
describe possible linear coupling of the quasimodes via the Doppler shift in
the frequency dependence of the linear gain. A similar coupling in the
coordinate space arises in the systems with inhomogeneous refractive index or
pump power distribution~\cite{deyc05a,zait10}. Below we argue, however, that the
off-diagonal elements $V_{kk'} (\omega_l)$ are negligible, i.e., the quasimode
wavefunctions correctly describe the lasing modes in the linear approximation. 

First, we evaluate the velocity average in Eq.~\eq{Vlkk}. In the Doppler limit 
\eqn{
  |\omega_l - \nu| \sim \nu \wbar v / c \gg \gamma_\perp,
\label{Dop}
}
typical for the gas lasers, the function~\eq{Dom} can be approximated~as
\eqn{
  D (\omega) \simeq \gamma_\perp \l[ \pi \delta (\Delta \omega) + i \msf P
  \frac 1 {\Delta \omega} \r], \quad \Delta \omega \equiv \omega - \nu,
\label{delta}
}
where $\msf P$ denotes the principal value. Then the angular average is given~by
\aln{
  &\l< D (\omega - q v \cos \varphi) \r>_\varphi \notag \\
  &= \gamma_\perp \l[ \frac {\theta \l(qv - |\Delta \omega| \r)} {\sqrt{q^2 v^2
  -\Delta \omega^2}}  +  i \frac {\theta \l(|\Delta \omega| - qv\r)}
  {\sqrt{\Delta \omega^2 - q^2 v^2}} \r]\! ,
\label{D_ang}
}
where $\theta (x)$ is the Heaviside step function. After the averaging over the
two-dimensional Maxwell distribution,
\eqn{
  W(v)\, dv = \frac 1 {\pi \wbar v^2} e^{- v^2/ \wbar v^2} v\, dv,
}
we find 
\aln{
  \l<  D (\omega - \mbf q \cdot \mbf v) \r>_{\mbf v} = \frac {\gamma_\perp} {q
  \wbar v} \l[ \frac 1 {2 \sqrt \pi} e^{ - \l(\Delta \omega/q
  \wbar v \r)^2 } \!
  + \frac i \pi F \l(\frac {\Delta \omega} {q \wbar v} \r) \r],
\label{D_av}
}
where $F(x) \equiv \exp (- x^2) \int_0^x \exp (y^2)\, dy$ is the Dawson
integral~\cite{abra72}. As a function of~$q$, $\l<  D (\omega - \mbf q \cdot
\mbf v) \r>_{\mbf v}$ changes on the scale $\Delta q \sim |\Delta \omega| /
\wbar v \sim \nu /c$, for typical~$\Delta \omega$.

Next, we note that $\psi_{k \mbf q}$ and $\phi_{k \mbf q}$ are significant only
when the magnitude of $\mbf q$ lies within the band $|q - q_k| \lesssim \delta
q_k$, where $q_k = \sqrt \epsilon\, \text{Re}\, \Omega_k / c$ and $\delta q_k =
- \sqrt \epsilon\, \text{Im}\, \Omega_k / c$. Therefore, the matrix elements
$V_{kk'} (\omega_l)$ vanish if $|q_k - q_{k'}| \gg \delta q_k$. In the reverse
case, $|q_k - q_{k'}| \lesssim \delta q_k$, we can pull $\l<  D (\omega_l - \mbf
q \cdot \mbf v) \r>_{\mbf v}$ out of the integral~\eq{Vlkk}, since this function
is constant on the scale $\delta q_k \ll \Delta q$. Then Eq.~\eq{biorthq} yields
the diagonality property, 
\eqn{
  V_{kk'} (\omega_l) = \l<  D (\omega_l - \mbf q \cdot \mbf v) \r>_{\mbf v, q =
  q_0} \delta_{kk'},
\label{Vlkk_diag}
}
where the substitution of $q = q_0 \equiv \sqrt \epsilon\, \nu / c$ is
consistent with the level of approximation in Eq.~\eq{mod_ampl}. We note that
the linear-gain frequency dependence, given by the real part of Eq.~\eq{D_av},
has a Gaussian shape, similar to one-dimensional gas lasers [see Eq.~\eq{L_av}].

\subsection{Nonlinear terms}
\label{sec:nlin}

The resonant dependence of the wavefunctions $\psi_{k \mbf q}$ and $\phi_{k \mbf
q}$ on~$q$, mentioned previously, makes it possible to simplify the nonlinear
terms of Eq.~\eq{mod_ampl} as well. The condition $|q - q_k| \lesssim \delta
q_k$ implies one of the following restrictions on the integration domain in
Eq.~\eq{Plk3}: (i)~$|\mbf q' - \mbf q''| \lesssim \delta q_k$, (ii)~$|\mbf q -
\mbf q''| \lesssim \delta q_k$, or (iii)~$|\mbf q + \mbf q'| \lesssim \delta
q_k$. Only in the first case we can estimate $D_\parallel [(\mbf q' - \mbf q'')
\cdot \mbf v] \sim \min(1, \gamma_\parallel/ \delta q_k \wbar v)$. In the cases
(ii) and (iii), we have $D_\parallel [(\mbf q' - \mbf q'') \cdot \mbf v] \sim
\gamma_\parallel c /\nu \wbar v \ll \min(1, \gamma_\parallel/ \delta q_k \wbar
v)$. Similarly, $D_\parallel [\omega_l - \omega_{l'} + (\mbf q' - \mbf q'')
\cdot \mbf v] \ll \min(1, \gamma_\parallel/ \delta q_k \wbar v)$ in all three
cases. Thus, we can neglect the terms in Eq.~\eq{Plk3} that are proportional to
$(1 - \delta_{ll'})$ and rewrite this equation with the new integration
variables $\mbf Q = (\mbf q' + \mbf q'')/2$ and $\delta \mbf q = \mbf q' - \mbf
q''$ corresponding to the choice~(i):
\aln{
  \mc P_{lk}^{(3)} = &i \nu^2 c^{(3)} \sum_{l' = 1}^{N_{\text m}} \frac 1 {(2
  \pi)^2} \l< \int_0^{2 \pi} d \varphi \int_0^{2 \pi} d \varphi'\, Y (\varphi,
  \varphi', \mbf v)\r. \notag \\   
  &\times\l. \vphantom{\int_0^{2 \pi}}
   D (\omega_l - q_0 v \cos \varphi)\, \mc L (\omega_{l'} - q_0 v
  \cos \varphi')  \r>_{\mbf v},
\label{Plk3res}
}
where $\mc L (\omega) \equiv \text{Re} D (\omega)$, and $q = Q = q_0$ is set in
all nonresonant functions. The value of the correlator
\aln{
  Y (\varphi, \varphi', \mbf v) = &\frac {q_0^2} {(2 \pi)^4} \int_0^\infty \! dq
  \int_0^\infty \! dQ \int d(\delta \mbf q)\, D_\parallel (\delta \mbf q \cdot
  \mbf v) \notag \\  
  &\times \phi_{k \mbf q}^* E_{l, \mbf q + \delta \mbf q} E_{l', \mbf Q + \delta
  \mbf q/2}^* E_{l', \mbf Q - \delta \mbf q/2} 
\label{Y}
}
is system specific. Here $\varphi$ ($\varphi'$) is the azimuthal angle of the
vector $\mbf q$ ($\mbf Q$) counted from the direction of~$\mbf v$. We can
approximate $D_\parallel (\delta \mbf q \cdot \mbf v) \approx 1$ under the
assumption of long living quasimodes, such that $\delta q_k \ll
\gamma_\parallel/ \wbar v$. In this case, the correlator depends only on the
direction of~$\mbf v$, but not on its magnitude. The absence of functions
$D_\parallel$ from the laser equations is an indication that the approximation
of constant population inversion is employed.

\section{Chaotic gas laser}

\subsection{Nonlinear terms in the laser equations}

We apply the general theory to a gas laser with wave-chaotic resonator. The
notion of a chaotic system implies that the shape of the resonator does not
permit a separation of variables in the wave equation and the relevant modes
have a typical wavelength~$\lambda = 2 \pi/ q_0 \ll L$. Famous examples of
chaotic resonators are the Sinai and stadium billiards and their desymmetrized
versions~\cite{reic04}. The classical ray dynamics, defined in the
short-wavelength limit, is assumed to be ergodic, i.e., the generic ray
trajectories fill densely the available phase space.  This assumption sets a
limitation on the degree of openness of the resonator: the lifetimes of the CF
modes contributing to the lasing modes are to be longer than the time needed
for the ray trajectories to establish ergodicity. This time is of the order of
inverse Lyapunov exponent. We will consider a weakly open resonator, such that
its eigenmodes are well approximated by real eigenfunctions $\psi_k (\mbf r)
\approx \phi_k (\mbf r)$ independent of the parameter~$\omega$. 

Let us examine a product of two wavefunctions evaluated at nearby points in the
wavevector space:
\aln{
  &\psi_{k, \mbf Q + \delta \mbf q/2}^*\, \psi_{k'\!, \mbf Q - \delta \mbf q/2} 
  = \int d (\delta \mbf r)\, e^{-i \mbf Q \cdot \delta \mbf r} \! \int_{\mc I} d
  \mbf R\, e^{i \delta \mbf q \cdot \mbf R} \notag \\
  &\times \psi_k \l(\mbf R - \frac {\delta \mbf r} 2 \r) \psi_{k'} \l(\mbf R +
  \frac {\delta \mbf r} 2 \r).
\label{corr}
}
Since the relevant magnitude of $\delta \mbf q$ is much smaller than the
optical wavenumber~$q_0$, the function $e^{i \delta \mbf q \cdot \mbf
R}$~changes on the length scale much longer than~$\lambda$. On the other hand,
the wavefunction product self-averages in the chaotic cavity over a correlation
area of order~$\lambda^2$~\cite{berr77}. Consequently, the product can be
substituted with its average value, expressed in terms of the correlation
function 
\eqn{
  C_{kk'} (\delta \mbf r) = A \l<\psi_k \l(\mbf R - \frac {\delta \mbf
  r} 2 \r) \psi_{k'} \l(\mbf R + \frac {\delta \mbf r} 2 \r) \r>_{\mbf R },
}
where the wavefunctions are assumed to be normalized. Different eigenmodes, $k
\ne k'$, are uncorrelated in the chaotic system, whereas the autocorrelation
function in two dimensions is approximated by a Bessel function~\cite{berr77},
\eqn{
  C_{kk} (\delta r) = J_0 (q_k \delta r).
}
This expression shows that the statistical characteristics of the eigenstates
are isotropic (this approximation fails within a wavelength of the cavity
boundary). Scars~\cite{kapl98}, i.e., the excess probability density around
unstable short periodic ray trajectories, should not significantly influence
the correlation function. Indeed, the excess probability density depends on the
inverse Lyapunov exponent, which scales with the system size. On the other
hand, the fraction of the phase space occupied by a scar reduces with the
wavelength, which, in gas lasers, is several orders of magnitude smaller than
the system size. The product~\eq{corr} is proportional to the Fourier transform
of the autocorrelation function, $(2 \pi/ q_k)\, \delta (Q - q_k)$, which makes
\eqn{
  \psi_{k, \mbf Q + \delta \mbf q/2}^*\, \psi_{k'\!, \mbf Q - \delta \mbf q/2} 
  = \frac {(2 \pi)^3} {q_k A}\, \delta_{kk'}\, \delta (Q - q_k)\, \delta_{\mc I}
  (\delta \mbf q).
}

We use this result to calculate the third-order contribution~\eq{Plk3res} to the
laser equations. An important simplification that comes with the stochasticity
of the wavefunctions is the isotropy of the correlator~\eq{Y}. Hence, this
function can be pulled out of the angular integrals, and we find
\aln{
  &\mc P_{lk}^{(3)} = \frac {i \nu^2} {\epsilon^{3/2} A}\, c^{(3)} a_{lk}
  \sum_{l' = 1}^{N_{\text m}} \l( \sum_{k'} |a_{l' k'}|^2 \r) \notag \\
  &\times \l< \l< D (\omega_l - q_0 v \cos \varphi)
  \r>_\varphi \l< \mc L (\omega_{l'} - q_0 v \cos \varphi) \r>_\varphi \r>_v.
\label{Plk3ch}
}

\subsection{Single-mode regime}
\label{sec:sing}

We apply the laser equations~\eq{mod_ampl}, with the right-hand side given by
Eqs.~\eq{Plk1exp}, \eq{Vlkk_diag}, and~\eq{Plk3ch}, to determine the dependence
of the mode intensity on the frequency in the single-mode regime. If the
pumping level is not far above the threshold, only one quasimode will
contribute to the lasing mode. Consequently, a single term with $l' = l$ and
$k' = k$ remains in the summations in Eq.~\eq{Plk3ch}. Taking the imaginary
part of Eq.~\eq{mod_ampl}, we obtain an expression for the
intensity, $I_k = |a_{lk}|^2/ \epsilon$,
\eqn{
  \frac {I_k (\omega)} {I_0} = \frac {\l< \l< \mc L (\nu - q_0 v \cos \varphi)
  \r>_\varphi^2 \r>_v} {\l< \l< \mc L (\omega - q_0 v \cos \varphi)
  \r>_\varphi^2 \r>_v} \l[ e^{- \l(\frac {\Delta \omega} {q_0
  \wbar v} \r)^2 }\! - \frac {\Delta N_{0,k}} {\Delta N_0} \r],
\label{I_om}
}
where the intensity unit
\eqn{
  I_0 = \frac A {8 \sqrt \pi} \frac {\hbar^2 \gamma_\perp \gamma_\parallel}
  {d^2 q_0 \wbar v} \l< \l< \mc L (\nu - q_0 v \cos \varphi) \r>_\varphi^2
  \r>_v^{-1}
}
is defined by the condition $I_k (\nu) \to I_0$, as $\Delta N_0 \to \infty$.
The threshold value of the pumping strength $\Delta N_0$ at $\omega = \nu$ is
given by the parameter
\eqn{
  \Delta N_{0,k} = \kappa_k \frac {\epsilon A}{\sqrt \pi \mc N} \frac {\hbar
  q_0 \wbar v} {d^2 \nu},
}
where $\kappa_k = - \text{Im}\, \Omega_k$ is the decay rate of the quasimode. 

When evaluating~$\l< \l< \mc L \r>_\varphi^2 \r>_v$, it is not possible to use
the approximation~\eq{D_ang} for the angular average, because the velocity
integral would diverge. The angular integral can be calculated analytically for
a finite value of $\gamma_\perp / q_0 \wbar v$, however, the resulting
expression is rather cumbersome. The remaining integral over the
distribution~$W(v)$ has to be evaluated numerically. In the special case $\omega
= \nu$ the result can be expressed in terms of a special function:
\aln{
  &\l< \l< \mc L (\nu - q_0 v \cos \varphi) \r>_\varphi^2 \r>_v \!= \frac x {2
  \pi}\, e^x\, \text{E}_1 (x), \\ 
  &x \equiv \l( \frac {\gamma_\perp} {q_0 \wbar v} \r)^2 \!,
}
where $\text{E}_1 (x) = \int_x^\infty\! dy\, e^{-y} / y$~\cite{abra72}. 

\begin{figure}
\begin{center}
  \includegraphics[width = .6 \linewidth]{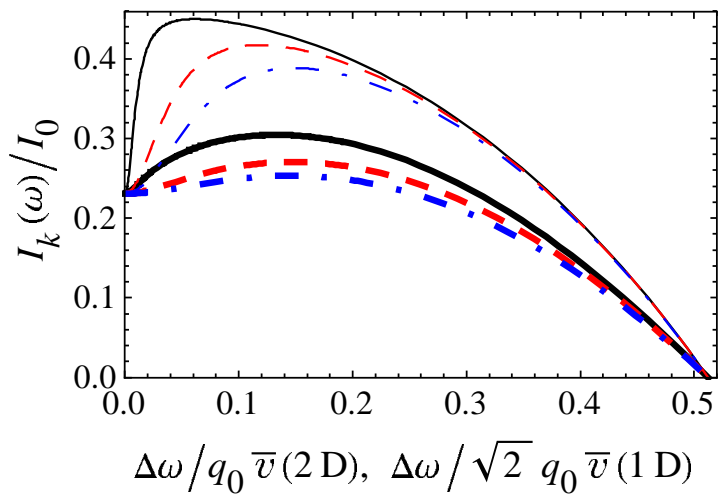}
\end{center}
  \caption{Relative intensity of the single lasing mode for the
  two-dimensional~(2D) chaotic laser [Eq.~\eq{I_om}, thick curves] and a
  one-dimensional~(1D) laser [Eq.~\eq{Iom1D}, thin curves] as a function of the
  frequency separation~$\Delta \omega$ from the atomic-line center. The scaling
  of $\Delta \omega$ for these systems differs by a factor of~$\sqrt 2$ (see
  the text). Parameters: relative pumping strength $\Delta N_0 / \Delta N_{0,k}
  = 1.3$; relative homogeneous linewidth $\gamma_\perp / q_0 \wbar v = 0.01$
  (solid curves), $0.04$ (dashed curves), and $0.07$ (dash-dotted curves). 
\label{fig1}}
\end{figure}

The relative intensity $I_k (\omega) / I_0$ is plotted in Fig.~\ref{fig1} for
several values of~$\gamma_\perp / q_0 \wbar v $ in the Doppler limit~\eq{Dop}.
The frequency dependence is characterized by a local minimum of intensity at the
center of the gain curve, $\omega = \nu$. The frequency interval where the
intensity is reduced has a width of the order of~$q_0 \wbar v$. Formally, this
follows from the analysis of the denominator of Eq.~\eq{I_om}. In fact, the
integral
\eqn{
  \l< \mc L (\omega - q_0 v \cos \varphi) \r>_\varphi = \frac 1 \pi \int_0^\pi
  \! \frac {d \varphi} {1 + (\Delta \omega - q_0 v \cos \varphi)^2/
  \gamma_\perp^2}
}
receives its major contribution from the angles in the interval $\delta \varphi
\sim \gamma_\perp / q_0 v$ around $\varphi = \cos^{-1} \l( \Delta \omega / q_0
v\r)$. Therefore, the integral changes its order of magnitude from $\gamma_\perp
/ q_0 v$, for $|\Delta \omega| < q_0 v - \gamma_\perp$, to $(\gamma_\perp /
\Delta \omega)^2$, for $|\Delta \omega| > q_0 v + \gamma_\perp$, i.e., $q_0
v$~is the characteristic frequency scale. 

The results for the chaotic laser are compared in Fig.~\ref{fig1} with the
relative mode intensity in a one-dimensional gas laser (\ref{sec:1Dgl}). In the
former (latter) case, $\Delta \omega$~is normalized by $q_0 \wbar v$ ($\sqrt 2
q_0 \wbar v$). Apart from the visual convenience, the difference in scaling is
motivated by the relation, $\wbar v_{\text{2D}} = \sqrt 2\, \wbar
v_{\text{1D}}$, between the root mean square velocities in two and one
dimensions at a given temperature. Lasers with uniaxial resonators exhibit a
much narrower intensity minimum, known as Lamb dip~\cite{sarg74}. Its width is
of the order of the homogeneous linewidth $\gamma_\perp \ll q_0 \wbar v$. In
order to understand the origin of the dip, it is convenient to treat the lasing
mode as a superposition of left- and right-moving waves. The atoms moving with
the velocity $v_x \approx \Delta \omega /q_x$ effectively interact with only
one of the two waves, depending on the sign of $\Delta \omega$ and~$v_x$.
However, when $|\Delta \omega| \lesssim \gamma_\perp$, the atoms at rest
interact with both running waves. This leads to a stronger gain saturation near
$\Delta \omega = 0$ than away from the line center. In the two-dimensional
chaotic laser this effect is washed out, since a mode with the frequency
$\omega$ draws its gain from a larger ensemble of atoms with velocities $v >
|\Delta \omega| / q_0$.  In particular, let us model the chaotic wavefunction
locally by a random superposition of partial plane waves with wavevectors of
fixed magnitude $\sqrt \epsilon\, \omega / c$ and isotropically distributed
directions~\cite{berr83}. Then, all partial waves of the central mode $\omega =
\nu$ shall interact with the atoms at rest, but each partial wave interacts, in
addition, with atoms moving perpendicular to its direction of propagation.
Thus, the minimum of intensity results from a more subtle interplay of linear
amplification and nonlinear saturation on the same frequency scale. 

The results obtained above for a wave-chaotic resonator cannot be directly
applied to resonators with mixed or regular classical ray dynamics, where the
hypothesis of random, isotropically distributed, and uncorrelated eigenstates is
no longer valid. Nevertheless, one can conjecture that increasing dimensionality
of the system will make the intensity minimum shallower and wider (its depth and
width being defined relatively to the intensity maximum and the inhomogeneous
linewidth, respectively), but will not lead to its complete disappearance. As a
practical consequence, the frequency stabilization~\cite{free70} and
spectroscopy~\cite{szok66} might still be possible in imperfect
quasi-one-dimensional resonators, where the lasing mode deviates from a
superposition of two counter-propagating plane waves. It is worth mentioning
that in both uniaxial and chaotic resonators the relative depth of the minimum
decreases with increasing~$\gamma_\perp$.

\section{Conclusions}

In the present article, the semiclassical laser theory was extended to gas
lasers with two-dimensional resonators of arbitrary shape. It was shown that
the linear coupling between the modes, resulting from the Doppler shift in the
gain frequency dependence, is negligible. The nonlinear coupling was considered
in the third order of the perturbation theory in the electric field. The
nonlinear terms in the laser equations that arise due to pulsations of the
population inversion were identified and neglected. The criterion $(\nu/c)\,
\wbar v \gg \gamma_\parallel$, allowing one to approximate the population
inversion as constant, is usually well fulfilled ($\nu$~is the
atomic-transition frequency and $\wbar v$~is the average velocity). 

The general theory was applied to lasers with two-dimensional weakly open
resonators of wave-chaotic geometry. In the short-wavelength limit, the
eigenfunctions of the resonator can be locally approximated by a superposition
of plane waves with a fixed wavelength propagating in random directions. The
applicability condition for this description can be estimated as $c/\nu \ll w
\ll L$, where $L$~is the size of the resonator and $w$~is the width of the
output window. The isotropic statistical characteristics of the eigenfunctions
lead to decoupling of the field distribution from the Doppler-broadened gain
curve in the nonlinear terms of the laser equations. The single-mode intensity,
as a function of the laser frequency, has a local minimum at the frequency of
the atomic transition. The width of the minimum scales approximately with the 
inhomogeneous broadening $(\nu/c)\, \wbar v$ and only weakly depends on the
homogeneous linewidth~$\gamma_\perp$. This property distinguishes the intensity
minimum in chaotic resonators from the Lamb dip in uniaxial resonators, which
has the width of~$\gamma_\perp \ll (\nu/c)\, \wbar v$.

\ack

Financial support was provided by the Deutsche Forschungs\-gemein\-schaft via
the grant FOR557.

\appendix

\section{Constant-flux quasimodes}
\label{sec:cf}

We consider an open cavity with a real dielectric constant~$\epsilon (\mbf r) =
\text{ const}$ inside and $\epsilon = 1$ outside of the system's boundary. The
constant-flux (CF) modes~\cite{ture06} $\psi_{k \omega} (\mbf r)$, depending on
the parameter~$\omega > 0$, are defined as solutions of a non-Hermitian
eigenvalue problem. Explicitly, the functions must satisfy the differential
equation
\eqn{
  -c^2\, \nabla^2 \psi_{k \omega} = \omega^2  \psi_{k \omega}
\label{cfout}
}
in the \emph{exterior} of the cavity with the outgoing-wave boundary conditions
at infinity, while \emph{inside} the system the same mode satisfies a
different equation:
\eqn{
  -c^2\, \nabla^2 \psi_{k \omega} = \epsilon\, \Omega_k^2 (\omega) \psi_{k
  \omega}.
\label{cfin}
}
For any given value of~$\omega$, the complex eigenfrequency $\Omega_k (\omega)$
is quantized [$k$~is a discrete (multi-)index labeling the modes], because the
solutions are required to match smoothly at the interface. 

The conjugate wavefunctions $\phi_{k \omega} (\mbf r)$ obey Eq.~\eq{cfout} with
the incoming-wave boundary conditions outside the cavity and the equation
\eqn{
  -c^2\, \nabla^2 \phi_{k \omega} = \epsilon [\Omega_k^2 (\omega)]^* \phi_{k
  \omega}.
}
inside the system. The CF quasimodes and their conjugates are biorthogonal, and
can be chosen to satisfy the condition
\eqn{
  \int_{\mc I} d \mbf r\, \phi_{k \omega}^* (\mbf r)\, \psi_{k' \omega} (\mbf r)
  = \delta_{kk'},
\label{biorth}
}
where the integration is over the interior~$\mc I$. A similar relation,
\eqn{
  \frac 1 {(2 \pi)^2} \int d \mbf q\, \phi_{k \mbf q \omega}^* \psi_{k' \mbf q
  \omega} = \delta_{kk'},
\label{biorthq}
}
is valid in the $\mbf q$ representation, defined as in Eq.~\eq{Eq}.
Additionally, the wavefunctions possess the properties: $\phi_{k \omega} (\mbf
r) = [\psi_{k \omega} (\mbf r)]^*$ and $\phi_{k \mbf q \omega} = \psi_{k,- \mbf
q, \omega}^*$.

A Fourier component $E_{\omega} (\mbf r)$ of the lasing field can be expanded
in the CF modes~as
\aln{
  &E_{\omega} (\mbf r) = \epsilon^{-1/2} \sum_k a_{k \omega}\,
  \psi_{k \omega} (\mbf r),
\label{normmod} \\
  &a_{k \omega} = \epsilon^{1/2} \int_{\mc I} d \mbf r\, \phi_{k \omega}^* (\mbf
  r)\,  E_{\omega} (\mbf r).
\label{normamp}
}
When continued to the exterior, this expansion yields a wave at the
frequency~$\omega$ propagating in the free space away from the system. In the
stationary lasing regime $\omega$ becomes restricted to a finite number of
values corresponding to the frequencies of lasing modes. To guarantee the
realness of the field~$E (\mbf r, t)$, we need to extend the definition of the
CF quasimodes to negative frequencies~by
\eqn{
  \psi_{k, - \omega} (\mbf r) = \phi_{k \omega} (\mbf r), \quad \phi_{k, -
  \omega} (\mbf r) = \psi_{k \omega} (\mbf r).
\label{negom}
}
Since the eigenfunctions $\phi_{k \omega} (\mbf r)$ depend only on the squared
complex frequency~$\Omega_k^2 (\omega)$, the frequencies~$\Omega_k (\omega)$ are
determined up to an overall sign. To describe the field distribution with the
outgoing flux and no incoming flux, the sign is chosen according to the
requirements:
\aln{
  &\text{Re} [\Omega_k (\omega)] > 0, \quad \text{Im} [\Omega_k (\omega)] < 0
  \quad (\omega > 0), \\
  &\Omega_k (- \omega) = - [\Omega_k (\omega)]^*.
}
In connection with the definition~\eq{negom}, this choice provides for negative
$\text{Im} [\Omega_k (\omega)]$ for any~$\omega$, as is expected from the
decaying quasimodes.

\section{One-dimensional gas laser}
\label{sec:1Dgl}

We apply the general theory to a one-dimensional gas laser that was studied in
detail in Ref.~\cite{sarg74}. Specifically, we consider a resonator defined by
$\epsilon = 1$ in the region $0 < x < L$, with a perfect mirror at $x = 0$ and a
weakly transmitting mirror at $x = L$. The quasimode eigenfunctions have the
form
\eqn{
  \psi_k (x) = \frac {e^{i \Omega_k x / c}} {i \sqrt{2 L}} - \frac {e^{-i
  \Omega_k x / c}} {i \sqrt{2 L}} \equiv \psi_k^{(+)} (x) + \psi_k^{(-)} (x),
\label{qmx1D}
}
where we approximate $q_k \equiv \text{Re}\, \Omega_k /c \approx \pi k /L$ ($k
= 1, 2, \ldots$) and assume $\delta q_k \equiv - \text{Im}\, \Omega_k /c \ll 
q_k$. In principle, $\delta q_k$ can be determined by modeling the
semitransparent mirror as a dielectric layer with large dielectric constant and
solving Eqs.~\eq{cfout} and~\eq{cfin}, but the specific value of $\delta q_k$
is not important for our purposes. In the $q$ representation, the functions
\eqn{
  \psi_{kq}^{(\pm)} = \frac 1 {\sqrt{2 L}}\, \frac {1 - e^{ -i (q \mp q_k \pm
  i\, \delta q_k) L }} {q_k - i\, \delta q_k \mp q}
\label{qmq1D}
}
have a resonant dependence on~$q$, peaked at~$\pm q_k$ and having a
width of~$\delta q_k$. 

As in the two-dimensional case, the velocity average of the homogeneous gain
curve can be computed within the approximation~\eq{delta}. Averaging over the
one-dimensional Maxwell distribution yields
\eqn{
  \l<  \mc L (\omega - q v) \r>_v = \sqrt{\frac \pi 2}\, \frac {\gamma_\perp} {q
  \wbar v} \exp \l[- \frac 1 2 \l( \frac {\Delta \omega} {q \wbar v} \r)^2 \r].
\label{L_av}
}
The third-order contribution is obtained from Eq.~\eq{Plk3}, adapted to the
one-dimensional case. The electric field can be represented as a superposition
of the quasimode wavefunctions~\eq{qmq1D}. The velocity averages depend on $q$
on a scale larger than~$\delta q_k$, and can be pulled out of the integrals and
evaluated at the resonant values of~$q$. The integrals over the wavefunctions
have a quite simple form in the coordinate representation:
\aln{
  &\frac 1 {(2 \pi)^3} \int\! dq\, dq'\, dq''\, \psi_{k, -q}^{(\alpha)}\, 
  \psi_{k', q + q' - q''}^{(\alpha')}\, \l[\psi_{k'', q'}^{(\alpha'')} \r]^*
  \psi_{k''', q''}^{(\alpha''')} \notag \\
  &= \int_0^L \! dx \, \psi_k^{(\alpha)} (x)\,  \psi_{k'}^{(\alpha')} (x)\,
  \l[\psi_{k''}^{(\alpha'')} (x)\r]^*  \psi_{k'''}^{(\alpha''')} (x),
\label{4int}
}
where $\alpha, \alpha', \alpha'', \alpha''' = +, -$ and we took into account
that $\phi_k^* (x) = \psi_k (x)$.  The phase of the integrand vanishes when the
indices obey the resonance condition
\eqn{
  \alpha k + \alpha' k' - \alpha'' k'' + \alpha''' k''' = 0.
}
In the single-mode near-threshold regime, when only one quasimode is excited,
six of the 16 possible combinations of indices are resonant. However, only four
combinations, $(\alpha, \alpha', \alpha'', \alpha''') = (+-++)$, $(-+--)$,
$(+---)$, $(-+++)$, yield $q' = q''$ at resonance, which results in $D_\parallel
= 1$. In the limit $\delta q_k L \ll 1$, the resonant value of the
integral~\eq{4int} is equal to~$1/4L$. The velocity averages in Eq.~\eq{Plk3}
with the resonant values of $q = \pm q_k$ and $q' = q'' \pm q_k$  sum up to $4
\l< D (\omega - q_k v) \l[\mc L (\omega - q_k v) + \mc L (\omega + q_k v) \r]
\r>_v$. This quantity was calculated in Ref.~\cite{sarg74}. The real part in the
Doppler limit is given~by
\aln{
  &\l< \mc L (\omega - q v) \l[\mc L (\omega - q v) + \mc L (\omega + q v) \r]
  \r>_v \notag \\
  &= \sqrt{\frac \pi 8}\, \frac {\gamma_\perp} {q \wbar v} \exp \l[- \frac 1 2
  \l( \frac {\Delta \omega} {q \wbar v} \r)^2 \r] \l[ 1 + \mc L (\omega) \r].
}
The imaginary part of Eq.~\eq{mod_ampl} yields an expression for the mode
intensity $I_k = |a_{lk}|^2$:
\eqn{
  \frac {I_k (\omega)} {I_0} = 2\, \frac {1 - \frac {\Delta N_{0,k}} {\Delta
  N_0}\, e^{\frac 1 2 \l(\frac {\Delta \omega} {q_0 \wbar v} \r)^2 }} {1 + \mc L
  (\omega)},
\label{Iom1D}
}
where we approximated $q_k \approx q_0$. The parameters 
\aln{
  &I_0 = \frac {\hbar^2 \gamma_\perp \gamma_\parallel L} {2 d^2 \nu}, \\
  &\Delta N_{0,k} = \frac {\delta q_k\, \hbar \wbar v L} {\sqrt 2\, \pi^{3/2}\,
  d^2 \mc N}
}
are defined in Sec.~\ref{sec:sing}.

\section*{References}
\bibliography{laser}

\begin{thebibliography}{10}

\bibitem{sarg74}
M.~Sargent~III, M.~O. Scully, and W.~E. Lamb, Jr.
\newblock {\em Laser Physics}.
\newblock Addison-Wesley, Reading, 1974.

\bibitem{hake84}
H.~Haken.
\newblock {\em Laser Theory}.
\newblock Springer, Berlin, 1984.

\bibitem{lamb64}
W.~E. Lamb, Jr.
\newblock {\em Phys. Rev.}, 134:1429, 1964.

\bibitem{free70}
C.~Freed and A.~Javan.
\newblock {\em Appl. Phys. Lett.}, 17:53, 1970.

\bibitem{szok66}
A.~Sz\"oke and A.~Javan.
\newblock {\em Phys. Rev.}, 145:137, 1966.

\bibitem{cao99}
H.~Cao, Y.~G. Zhao, S.~T. Ho, E.~W. Seelig, Q.~H. Wang, and R.~P.~H. Chang.
\newblock {\em Phys. Rev. Lett.}, 82:2278, 1999.

\bibitem{pols01b}
R.~C. Polson, A.~Chipouline, and Z.~V. Vardeny.
\newblock {\em Advanced Materials}, 13:760, 2001.

\bibitem{misi98}
T.~S. Misirpashaev and C.~W.~J. Beenakker.
\newblock {\em Phys. Rev. A}, 57:2041, 1998.

\bibitem{patr00}
M.~Patra, H.~Schomerus, and C.~W.~J. Beenakker.
\newblock {\em Phys. Rev. A}, 61:023810, 2000.

\bibitem{vivi03}
C.~Viviescas and G.~Hackenbroich.
\newblock {\em Phys. Rev. A}, 67:013805, 2003.

\bibitem{deyc05a}
L.~I. Deych.
\newblock {\em Phys. Rev. Lett.}, 95:043902, 2005.

\bibitem{ture06}
H.~E. T{\"u}reci, A.~D. Stone, and B.~Collier.
\newblock {\em Phys. Rev. A}, 74:043822, 2006.

\bibitem{zait10b}
O.~Zaitsev and L.~Deych.
\newblock {\em Phys. Rev. A}, 81:023822, 2010.

\bibitem{zait10}
O.~Zaitsev and L.~Deych.
\newblock {\em J. Opt.}, 12:024001, 2010.

\bibitem{berr77}
M.~V. Berry.
\newblock {\em J. Phys. A}, 10:2083, 1977.

\bibitem{berr83}
M.~V. Berry.
\newblock In G.~Iooss, R.~Helleman, and R.~Stora, editors, {\em Chaotic
  Behaviour of Deterministic Systems}, page 171. North-Holland, Amsterdam,
  1983.

\bibitem{zait07}
O.~Zaitsev.
\newblock {\em Phys. Rev. A}, 76:043842, 2007.

\bibitem{ge08}
L.~Ge, R.~J. Tandy, A.~D. Stone, and H.~E. Tureci.
\newblock {\em Opt.\ Exp.}, 16:16895, 2008.

\bibitem{ture08}
H.~E. T{\"u}reci, L.~Ge, S.~Rotter, and A.~D. Stone.
\newblock {\em Science}, 320:643, 2008.
\newblock See also Supporting Online Material.

\bibitem{abra72}
M.~Abramowitz and I.~A. Stegun, editors.
\newblock {\em Handbook of Mathematical Functions}.
\newblock Dover, New York, 1972.

\bibitem{reic04}
L.~E. Reichl.
\newblock {\em The Transition to Chaos: Conservative Classical Systems and
  Quantum Manifestations}.
\newblock Springer, New York, 2004.

\bibitem{kapl98}
L.~Kaplan and E.~J. Heller.
\newblock {\em Ann.\ Phys.\ (N.Y.)}, 264:171, 1998.

\end{thebibliography}

\end{document}